\documentclass[a4paper,11pt]{article}
\pdfoutput=1 

\usepackage{jinstpub} 
\usepackage{graphicx}
\usepackage{mathtools}

\usepackage{subcaption}

\title{\boldmath gggg $x=1$}

\author[a,b,1]{Sridhar Tripathy,\note{Corresponding author.}}
\author[a,b,c]{Jaydeep Datta,}
\author[a,b,c,2]{Abhik Jash,\note{Present address: National Institute of Science Education and Research, Bhubaneswar, India.} }
\author[a,d,3]{Subhadip Bouri,\note{Summer project student.}}
\author[a,b]{Nayana Majumdar,}
\author[a,b]{Supratik Mukhopadhyay,}
\author[a,b]{Sandip Sarkar}

\affiliation[a]{Saha Institute of Nuclear Physics,\\Kolkata, India}
\affiliation[b]{Homi Bhabha National Institute,\\Mumbai, India}
\affiliation[c]{Bhabha Atomic Research Centre,\\Mumbai, India}
\affiliation[d]{Indian Institute of Technology,\\Kharagpur, India}
\emailAdd{sridhar.tripathy@saha.ac.in}

\abstract{We plan to build an imaging setup for material identification utilizing the Coulomb scattering of cosmic ray muons due to their interaction with the materials and tracking their trajectories with Resistive Plate Chambers (RPCs). To begin with, we consider a setup of six RPCs stacked in a parallel manner to read the position and timing information of the muons before and after their interaction with a phantom of a given material using a set of three RPCs for each phase. Here we present a simulation work carried out to study the image formation of phantoms of several materials. A detailed modeling of the imaging system consisting of six RPCs was done using GEANT4. Cosmic Ray Library (CRY) was used for generation of particles with the appropriate energy and zenith angle distribution. Three reconstruction algorithms were followed for material identification and image reconstruction, viz.~Point of Closest Approach (POCA), Iterative POCA and the Binned Cluster Algorithm. A weighted metric discriminator was calculated for target object identification. Using the algorithms, the imaging of the Region of Interest (ROI) lying between the two layers of RPCs was done. The time required to discriminate target objects and do the image reconstruction has been studied.}

\keywords{Muon Tomography, GEANT4, POCA algorithm, Resistive Plate Chambers}

\proceeding{XIV$^{\text{th}}$ Workshop on Resistive Plate Chambers And Related Detectors\\
  19-23 Feb, 2018\\
  Puerto Vallarta, Mexico}

\begin{document}
	\title{Material Identification with Cosmic Ray Muons using RPCs}
\maketitle
\flushbottom
\section{Introduction}
\label{sec:intro}
{
	
	Cosmic ray muons are leptons formed from the decay of high energy cosmic ray pions and kaons. They are highly lived (lifetime 2.2 $\mu$s.) charged particle with mass 105.6 MeV/c$^{2}$~\cite{PDG}, nearly 200 times mass of electron. Muons take part in week interaction as well as electromagnetic interaction. The cosmic muon flux is $\sim$ 10$^{4}$m$^{-2}$min$^{-1}$~\cite{PDG} at mean sea level. The distribution of zenith angle ($\theta$) of muons is proportional to cos$^{2}\theta$. Muons like other charged particles loose energy due to ionization. However mean momenta of muons at sea level is about 3$-$4 GeV/c~\cite{PDG}. Therefore being minimum ionization particle (MIP), they usually deposit minimal energy while passing through matter. Apart from ionization, muons take part in multiple Coulomb scattering (MCS) with atomic nuclei. The distribution of scattering angles due to MCS is considered as a Gaussian distribution~\cite{PDG} with mean at 0 and RMS width ($\sigma$) is given by 
	\begin{equation}
	\label{1}
	\Large
	\begin{aligned}
	\sigma= \frac {13.6}{\beta cp} \sqrt{\frac{L}{X_0}}\bigg(1+0.038\:ln\:{\frac{L}{X_0}}\bigg)
	\end{aligned}
	\end{equation}
	\begin{equation}
	\label{2}
	\Large
	\begin{aligned}
	{X_0}=\frac{716.4g/cm^2}{\rho} \frac{A}{Z(Z+1)ln\big(\frac{287}{\sqrt{Z}}\big)}
	\end{aligned}
	\end{equation}
	Where $\beta$ is the ratio between velocity of muon $v$ to velocity of light $c$. $p$ is the momentum of the muon, $X_{0}$ is the radiation length of the material, $L$ is the length of the material traversed. 
	$X_{0}$ is a material property and depends on density the of the material ($\rho$), the atomic mass ($A$) and the atomic number ($Z$). From equations ~\ref{1} and~\ref{2}, it can be seen that for high-$Z$ materials the radiation length is small and scattering angle distribution will have broader width. In dense and high-$Z$ materials, muons see larger nuclei and more in number, hence they scatter more than they scatter through the low-$Z$ materials. This phenomenon is the index of probing in muon scattering tomography (MST). This method; being a radiation hazard-free, freely available source can be a candidate for imaging of suspected cargo containers, nuclear waste imaging etc. This method can also be used along with the absorption tomography method for imaging of civil structures, interior of large monuments.

	We plan to develop a MST system consisting of 6 RPCs as muon tracking detectors. This system will be a small prototype setup, using RPCs of dimension 30 $\times$ 30 cm$^2$, which will be upgraded in the future for non-destructive tomography purposes. The present work carries out a GEANT4~\cite{GEANT4} simulation of the present setup to study the feasibility and scope of utilizing muon scattering in this direction.

}

\section{Simulation Environment}
\subsection{Geometry}
\label{21}
\begin{figure}[!tbph]
	\centering
	\includegraphics[width=.45\textwidth]{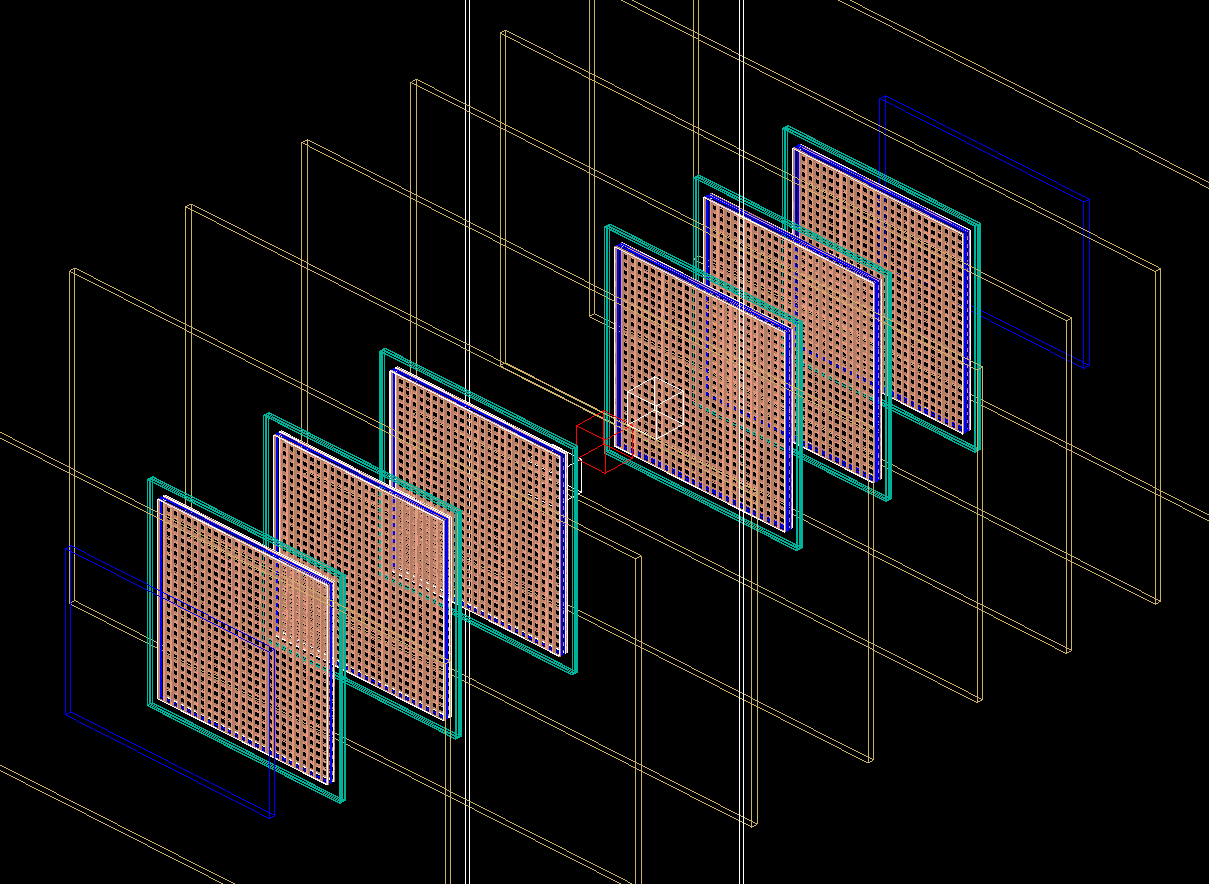}
	\caption{GEANT4 simulation of the setup.}
	\label{setup}	
\end{figure}

{The setup in GEANT4~\cite{GEANT4} simulation consisted of 6 RPCs of dimension 30 $\times$ 30 cm$^2$. The chambers are made up of a pair of bakelite plates of thickness 3 mm enclosing a gas gap of 2 mm. The outer surfaces of the plates were covered with a coating of graphite of thickness 500 microns to apply high voltages. The resistive electrode, thus prepared, were insulated by covering them with mylar foils of thickness 1mm. Signal pick-up panels made of copper strips of width 1 cm and pitch 1.2 cm were kept on the insulation layers in orthogonal directions. Copper, being a mid-Z material (Z$=$29) can do Coulomb scattering with the muons, hence causing slight deviation in their track. This fact was considered in the simulation to produce close to realistic scenario. A mixture of R$-$134A (a variant of Freon) and isobutane gases, in the ratio 95:5 was used as the gas mixture. The setup was placed inside a large concrete housing and the surrounding was filled with atmospheric air to mimic the lab environment. The simulation setup is shown in the figure~\ref{setup}. The RPCs were laid in the XY plane while the axis of the setup was along the Z- direction. Two plastic scintillator paddles were included for trigger purpose, with one placed on the top of the setup while the other at the bottom. 
The target objects were kept in a horizontal plane (parallel to the RPCs), inside the ROI lying between two sets of detectors. The volume of ROI in this case was 30 $\times$ 30 $\times$ 50 cm$^3$. Cubes of different materials of side 5 cm were chosen as targets. 
The upper set of RPCs were positioned closer to the plane of the phantoms than the lower ones. This type of arrangement allowed additional deviation in the muon tracks, which increased the upper bound of the spatial resolution of the detector. }

\vskip 1.6pt
\subsection{Particle Generator}
{ The cosmic ray library (CRY)~\cite{CR} was used as the particle generator for the simulation. It follows the Gaisser's parametrization~\cite{GZ} for cosmic ray muons at sea level with modification for low energies and higher zenith angles~\cite{CRR}. The cosmic ray events from a planar surface of area 60 $\times$ 60 cm$^2$ were generated. The latitude, altitude of the present experimental lab were implemented in the code. The cosmic ray flux is influenced by the solar activity~\cite{CRR}. Therefore a normal day other than solar maxima or minima was chosen, to avoid the dependence on the solar activity.}
\subsection{Tracking}
{The particle generation and tracking were governed by FTFP\_BERT physics list~\cite{FT}. This takes care of all the electromagnetic as well as weak interactions needed for the propagations of $e^-$, $e^+$, $\mu^-$, $\mu^+$ and $\gamma$. The particle hits on the sensitive detector volume were stored according to their event-IDs. The hits on the detector planes due to $\mu^-$, $\mu^+$ as well as $e^-$, $e^+$ and $\gamma$ were recorded. The storing of hits due to other particles except  $\mu^-$ and $\mu^+$ included the experimental noise that could occur due to secondary particles as well as background radiations etc. This can cause difficulty in track reconstruction. The hits on the upper set of detectors provided the incoming track while the lower set of detectors produced the outgoing track. In the simulation, from the event-ID information, the detector hits were associated with a particular muon path. The hits due to all the particles ($\mu^-$, $\mu^+$, \large$e^-$,\large$e^+$ and $\gamma$) are saved. These events leading to noisy tracks were rejected by applying two cuts. First, a track must have hits from all the detector layers. Second, a track must not have number of hits higher than the number detectors. Considering the energy spectrum of cosmic muons it was assumed that, they will deposit small amount of energy and won't undergo MCS inside the 2 mm gas volume}
\section{Reconstruction Algorithm}
{The reliability of the tomography method depends upon accurate estimate of interaction points of muons with the matter and precise idea of the amount of scattering. The reconstruction algorithms use the hit information from the detectors, momenta information of the incoming muons, and the energy deposition of the particles inside the detectors to reconstruct the tracks. The property of the tracks e.g. deviation (scattering), absorption, energy deposition etc. are exploited for image reconstruction. 
\vskip 4 pt		
In this work, the scattering phenomenon of the muons was investigated.  
Three reconstruction algorithms were probed for identifying the target objects placed in the ROI. All these algorithms are based on the assumption that the muons are scattered from a particular point and do not take the MCS into account. These found out the scattering angle and point of scattering for each of the event.
}
\subsection{Point of Closest Approach (POCA)}
{This is a geometrical algorithm which calculates the closest point of approach between two skew lines in 3D~\cite{POCA} or their meeting point which is taken as the scattering point. In these cases, the segment of the track through the ROI is considered as the POCA point. This is done by minimizing the distance, called the distance of closest approach ($D$), between these tracks. When the tracks do not intersect, the mid-point of these closest points are taken, which is called the POCA point.}
\subsection{Iterative Point of Closest Approach}	
{The POCA points do not necessarily fall inside the target object. For some of the tracks, $D$ becomes more than the size of the target. In this cited work~\cite{iPOCA}, another variant of POCA algorithm is described which uses an iterative method. If the tracks have $D$ less than a certain value (0.1 cm in this case), then the algorithm converges and returns the POCA point. But if $D$ is larger, two new points are chosen on the line joining the closest points at one-third and two-third locations of its length ($D$). Then new pair of tracks is constructed using the newly found points and again the length of $D$ is checked for the condition ($D$ < 0.1 cm). Iteration continues until the condition is satisfied.}
\subsection{Binned Cluster method}
{This algorithm~\cite{bristol}, like POCA, assumes that the incoming and outgoing tracks scatter from a common vertex point. The method defines an energy function $E$ for both XZ and YZ planes. The hits on the detector planes $h_{x_{i}}$, $h_{y_{i}}$ and the uncertainty $\sigma_{h_{x_{i}}}$, $\sigma_{h_{y_{i}}}$ in their measurement are given as the input. The $Z$- position $z_{i}$ of the detectors are taken from the simulation. The vertex coordinates $v_{x}$, $v_{y}$ and $v_{z}$ are the unknowns to be determined. The four slopes $k_{x,upper}$, $k_{x,lower}$, $k_{y,upper}$ and $k_{y,lower}$ are calculated from track fitting and are not taken as unknowns.
\begin{equation}
E_{x} = \sum_{i=1}^{3}\frac{(h_{x_{i}} - (v_{x} + k_{x,upper}\cdot t))^{2}}{\sigma^{2}_{h_{x_{i}}}}  +  \sum_{i=4}^{6}\frac{(h_{x_{i}} - (v_{x} + k_{x,lower}\cdot t))^{2}}{\sigma^{2}_{h_{x_{i}}}}
\end{equation}
where, t = z$_{i}$ - v$_{z}$. Similarly the energy function can be defined for the YZ plane and the total energy is given by:
\begin{equation}
\large
\label{ee}
\begin{aligned}
E=E_{x}+E_{y}
\end{aligned}
\end{equation}

The function E is minimized by the \textsl{minimize} function of MATLAB~\cite{Matlab} to obtain $v_{x}$, $v_{y}$ and $v_{z}$. The ROI can be divided into several volume segments. The vertices calculated in a sub volume are clustered depending upon the target material kept there. For each pair of vertices $v_{i}$, $v_{j}$ in a sub volume the metric distance $m_{ij}$ is calculated. For a particular material signature, the metric distance is then weighted by the product of scattering angles of those two tracks $\theta_{i}$ and $\theta_{j}$.

\begin{equation}
\label{ee}
\begin{aligned}
m_{ij} = |v_{i}-v_{j}|
\end{aligned}
\end{equation}

\begin{equation}
\label{ee}
\begin{aligned}
m_{ij} = \frac{|v_{i}-v_{j}|}{\theta_{i}\cdot \theta_{j}}
\end{aligned}
\end{equation} 
Therefore this weighting could be better with the momentum information of the tracks which was not implemented in this work. The ROI of dimension 30 $\times$ 30 $\times$ 50 cm${^3}$ was subdivided into 4 sub volumes of size 15 $\times$ 15 $\times$ 50 cm${^3}$ to contain the targets. The calculation of this weighted metric, depends on the number of events selected in a particular sub volume. Therefore the tracks with highest amount of scattering were sorted in descending order for each sub volumes and same number of events were chosen for each of them.}

\section{Results}
{		
\subsection{Image Reconstruction}{
Cosmic ray data of $\sim$ 17 hrs.~i.e. 3.5 $\times$ 10$^6$ muon events were generated for imaging the ROI of the present setup. For the reconstruction of the images, tracks having a scattering angle deviation 2 degrees (34 mrad) or more are selected. Three cubical blocks of side 5 cm of Pb, U and Ag are kept in the ROI on the same Z plane. 
\\ 
The 2D projection of reconstructed images using the POCA algorithm is shown in the figure~\ref{Pnores}. The inclusion of 1 cm spatial resolution, makes the algorithm fail to detect the target materials as shown in figure~\ref{PWires}.

\begin{figure}[!tbph]
	\centering
	\begin{minipage}[b]{0.48\textwidth}
		\includegraphics[width=\textwidth,height=0.83\textwidth]{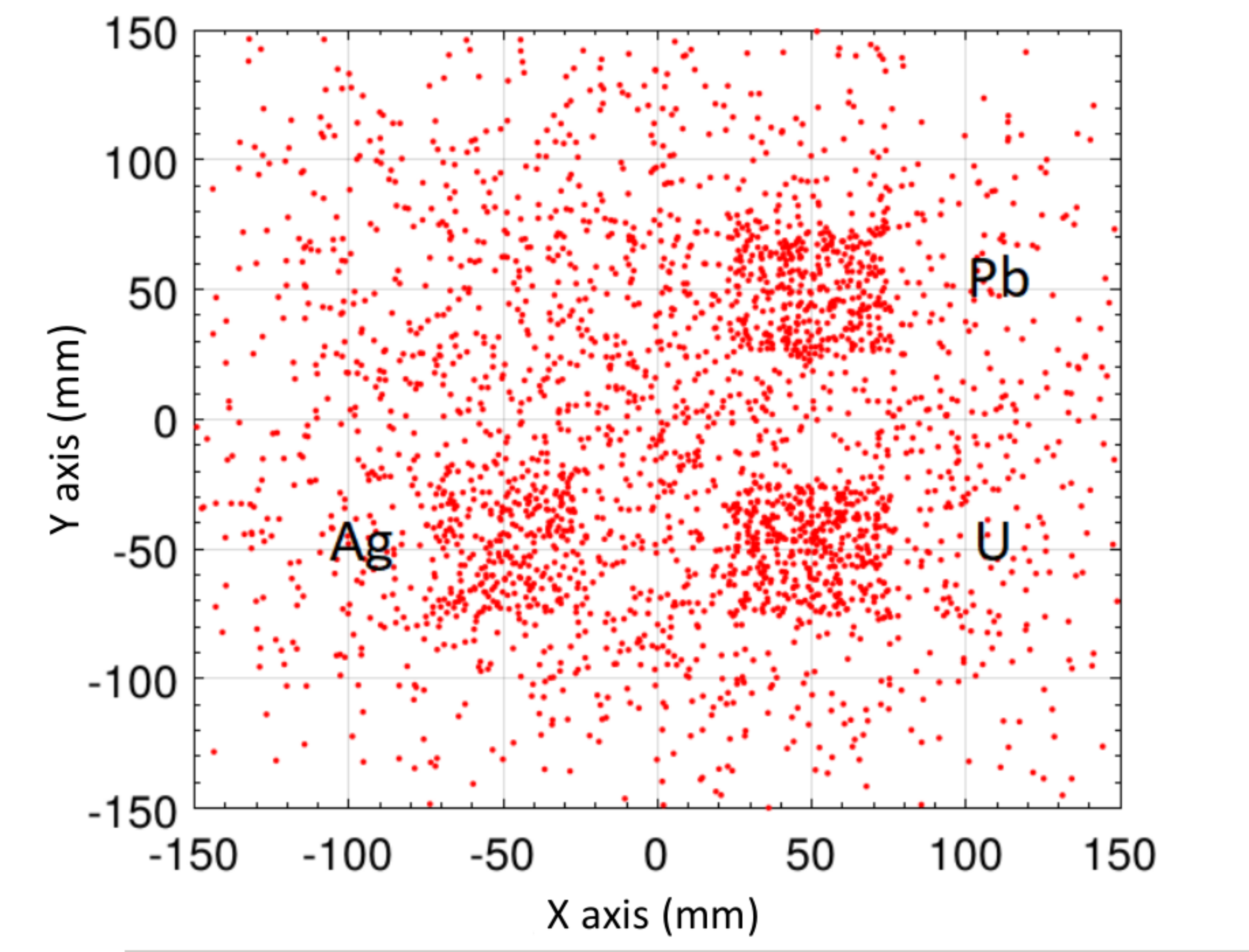}
		\caption{POCA reconstructed image (perfect resolution). }
		\label{Pnores}
	\end{minipage}
	\hspace*{0.1cm}
	\begin{minipage}[b]{0.48\textwidth}
		\includegraphics[width=\textwidth,height=0.83\textwidth,]{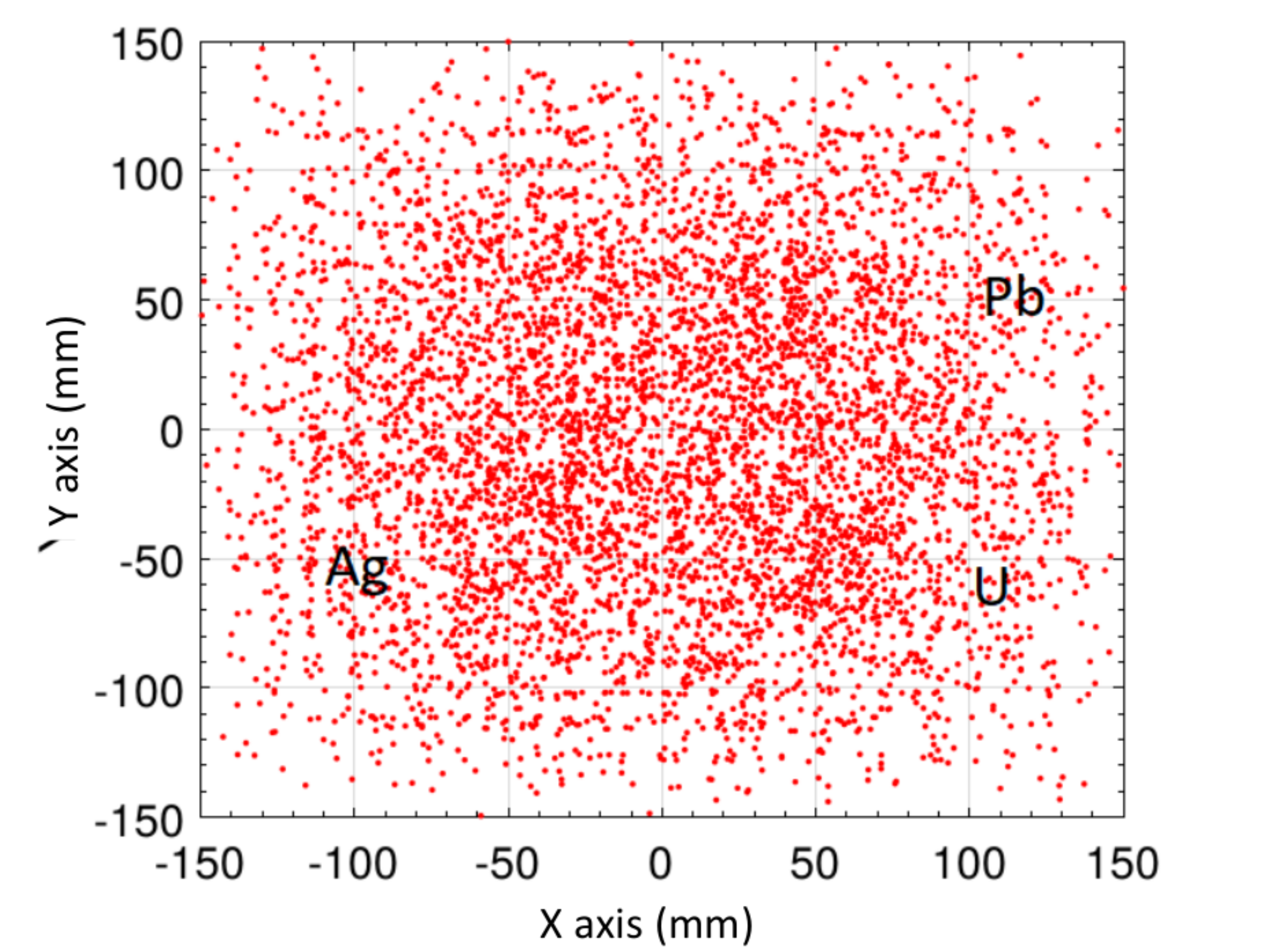}
		\caption{POCA reconstructed image (1 cm resolution).}
		\label{PWires}
	\end{minipage}
\end{figure}

Then the Iterative POCA algorithm is used for the same purpose. The figures ~\ref{iPnores}, ~\ref{iPWires} display the 2D projection of reconstructed images for the given set of target materials respectively.

\begin{figure}[!tbph]
	\centering
	\begin{minipage}[b]{0.48\textwidth}
		\includegraphics[width=\textwidth,height=0.83\textwidth]{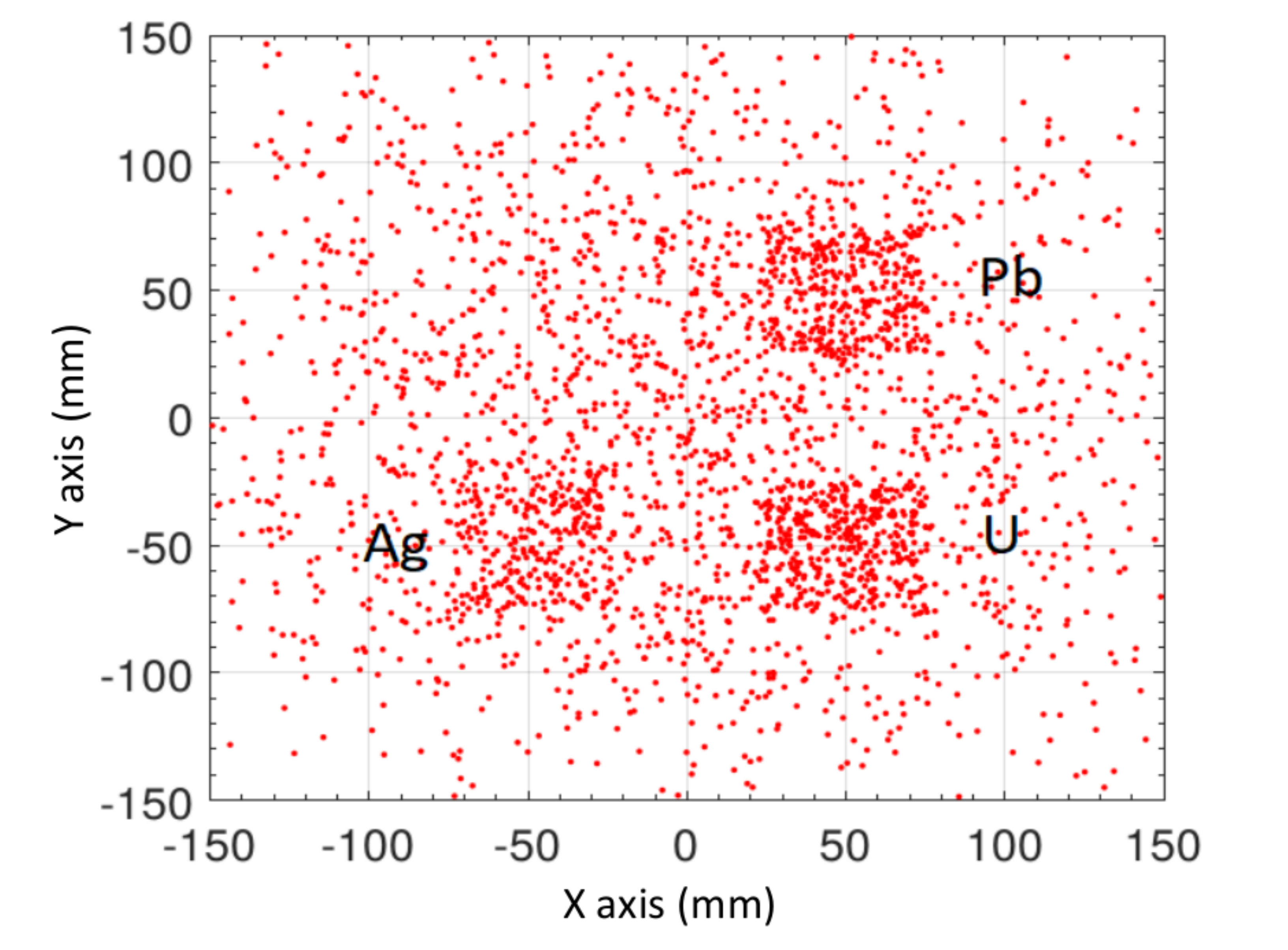}
		\caption{Iterative POCA reconstructed image (perfect resolution.) }
		\label{iPnores}
	\end{minipage}
	\hspace*{0.1cm}
	\begin{minipage}[b]{0.465\textwidth}
		\includegraphics[width=\textwidth,height=0.87\textwidth,]{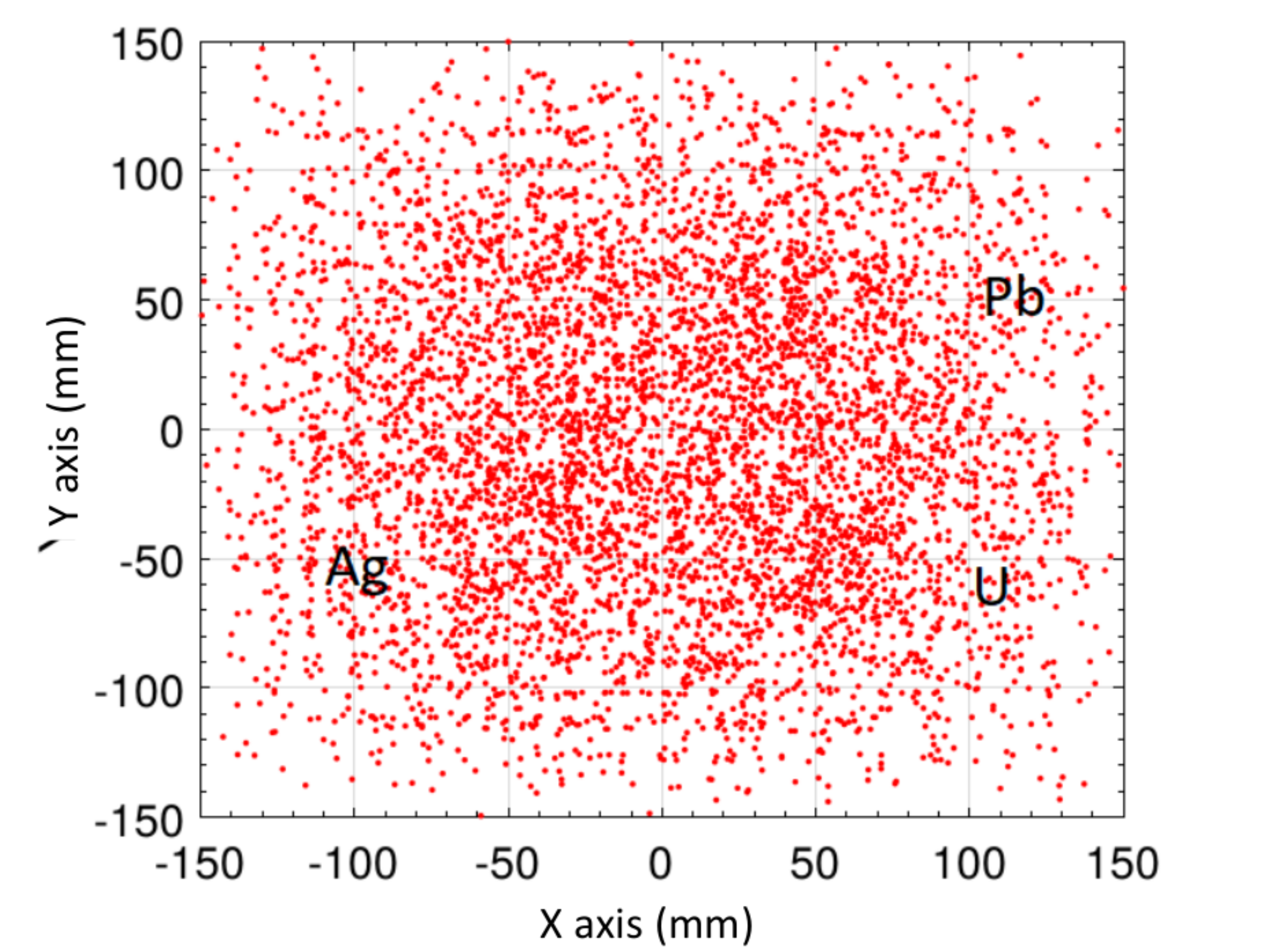}
		\caption{Iterative POCA reconstructed image (1 cm resolution).}
		\label{iPWires}
	\end{minipage}
\end{figure}

Then the bin cluster algorithm has been used for the same purpose. The vertices found after minimizing the function $E$, are supposed to be clustered where target materials are present. The figures~\ref{VPnores}, ~\ref{VPWires} display the 2D reconstructed images for the given set of target materials respectively. 

\begin{figure}[!tbph]
	\centering
	\begin{minipage}[b]{0.48\textwidth}
		\includegraphics[width=\textwidth,height=0.83\textwidth]{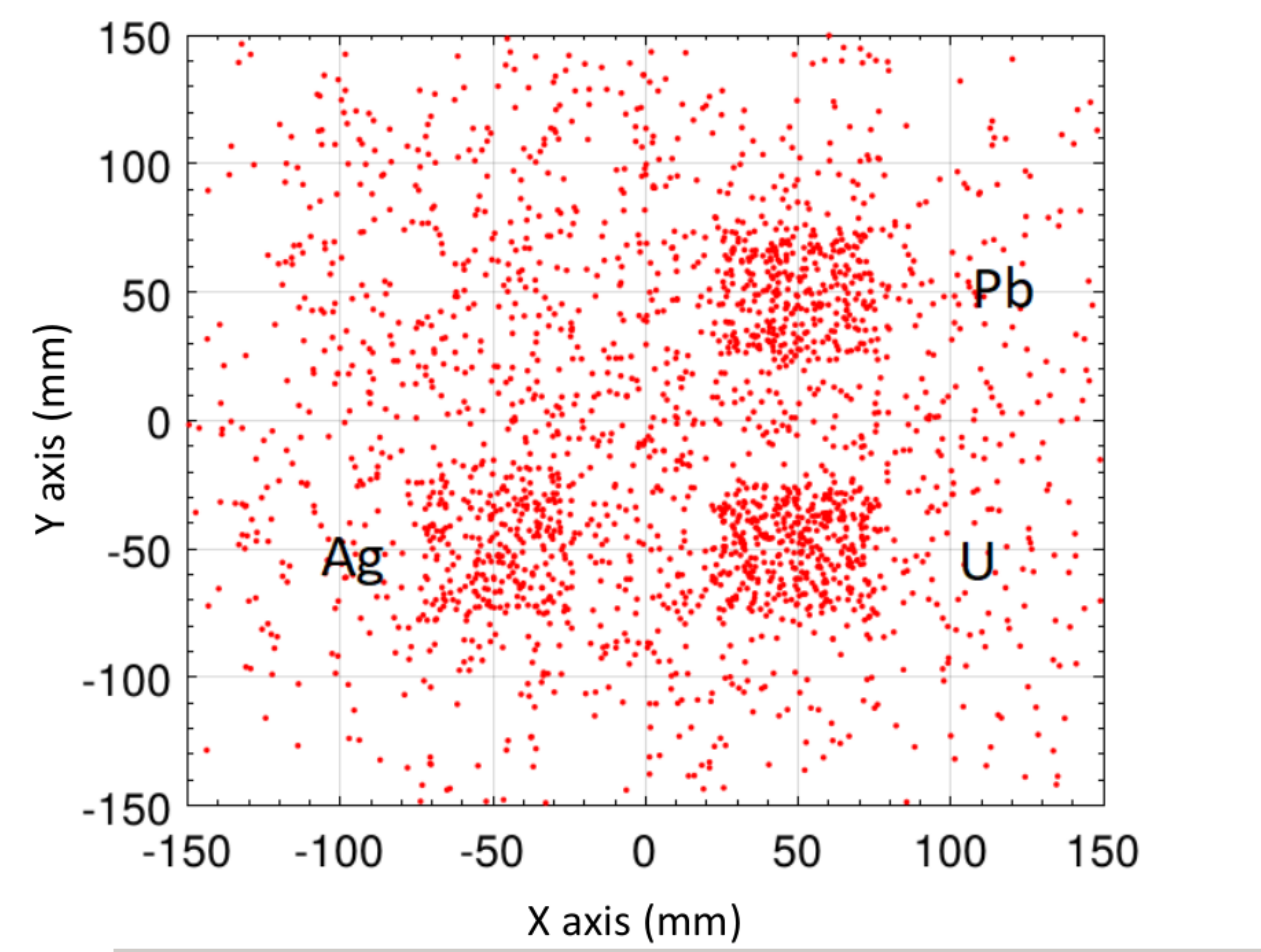}
		\caption{Binned cluster reconstructed image (perfect resolution). }
		\label{VPnores}
	\end{minipage}
	\hspace*{0.1cm}
	\begin{minipage}[b]{0.48\textwidth}
		\includegraphics[width=\textwidth,height=0.83\textwidth,]{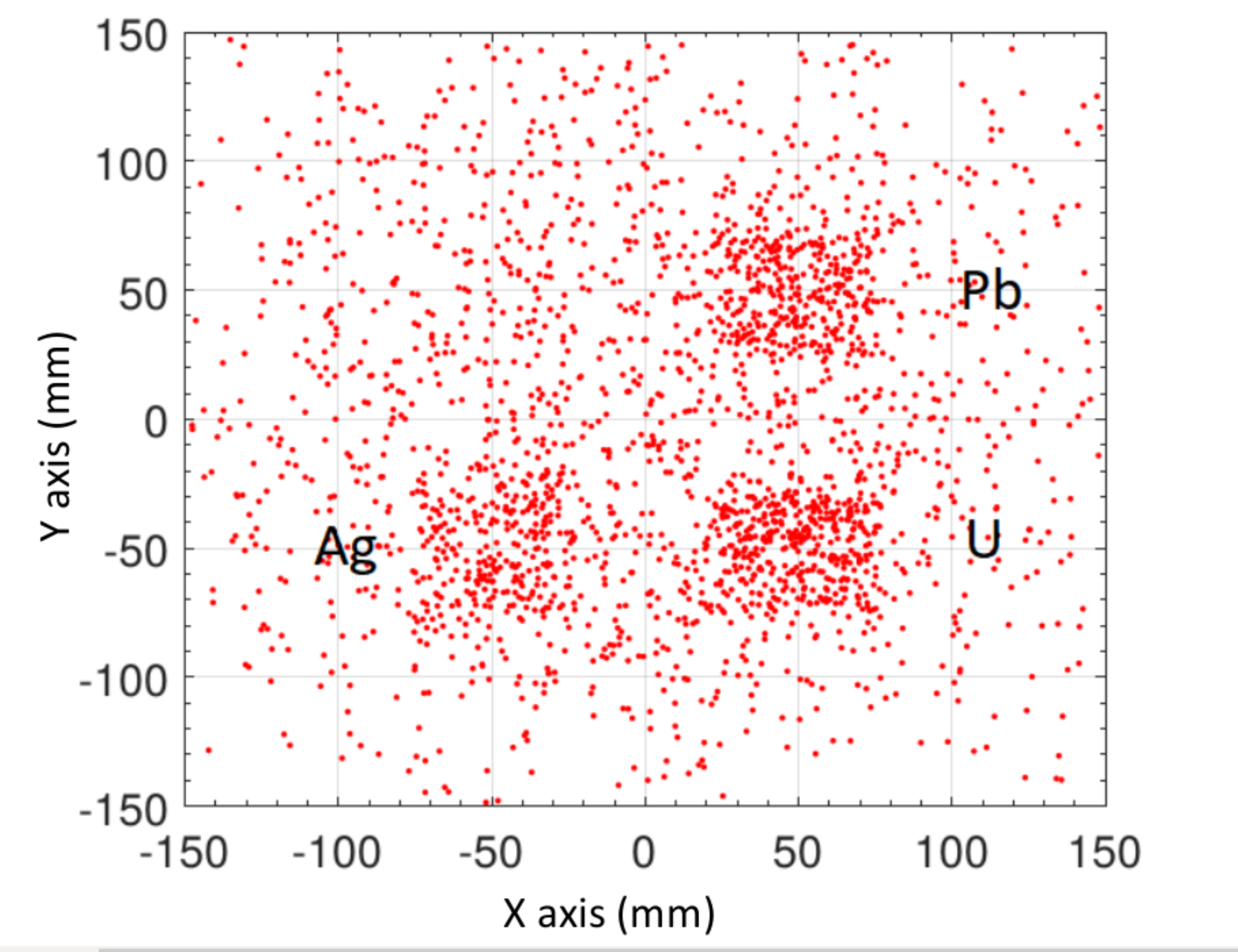}
		\caption{Binned Cluster reconstructed image (1 cm resolution).}
		\label{VPWires}
	\end{minipage}
\end{figure}
\vskip 3pt
It can be seen that for spatial resolution 1cm of the detector, the POCA and its variant, Iterative POCA are not able to discriminate the targets rather give a scatter output throughout the region of interest. Whereas the binned cluster algorithm is successful even when the detector has spatial resolution of 1 cm.
\vskip 3pt
\subsection{Identification of Materials using the Discriminator}
Further these algorithms were tested for a quicker material discrimination technique based on section~\ref{ee}. The comparison is done between these 3 algorithms for 1 hr. experimental data with 1 cm spatial resolution consideration.   

\begin{figure}[!tbph]
	\centering
	\begin{minipage}[b]{0.32\textwidth}
		\includegraphics[width=\textwidth,height=0.50\textheight,keepaspectratio,trim={1cm 7cm 1cm 7cm},clip]{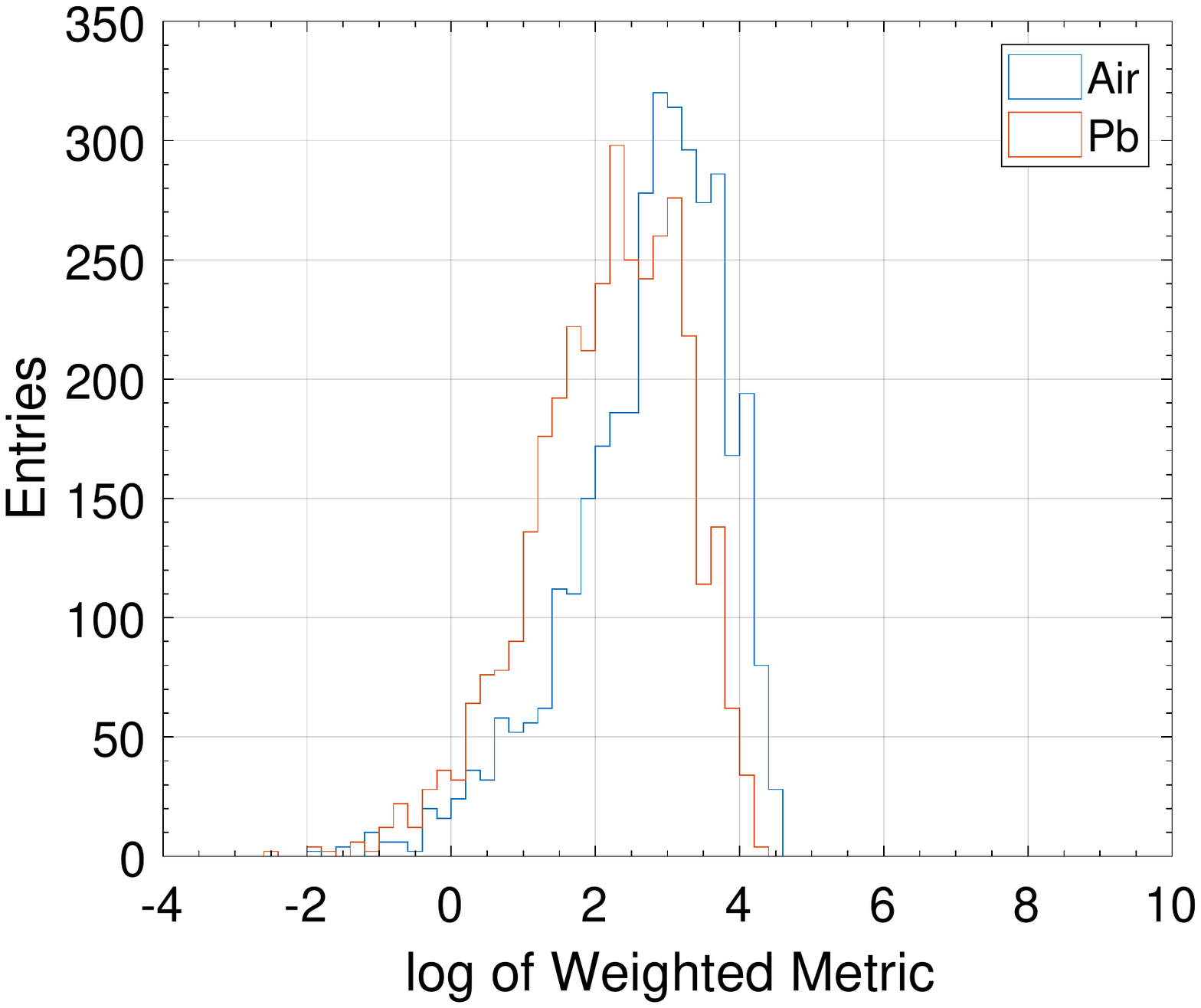}
		\caption{POCA.}
		\label{POCA}
	\end{minipage}
	\hspace*{0.05cm}
	\begin{minipage}[b]{0.32\textwidth}
		\includegraphics[width=\textwidth,height=0.50\textheight,keepaspectratio,trim={1cm 7cm 1cm 7cm},clip]{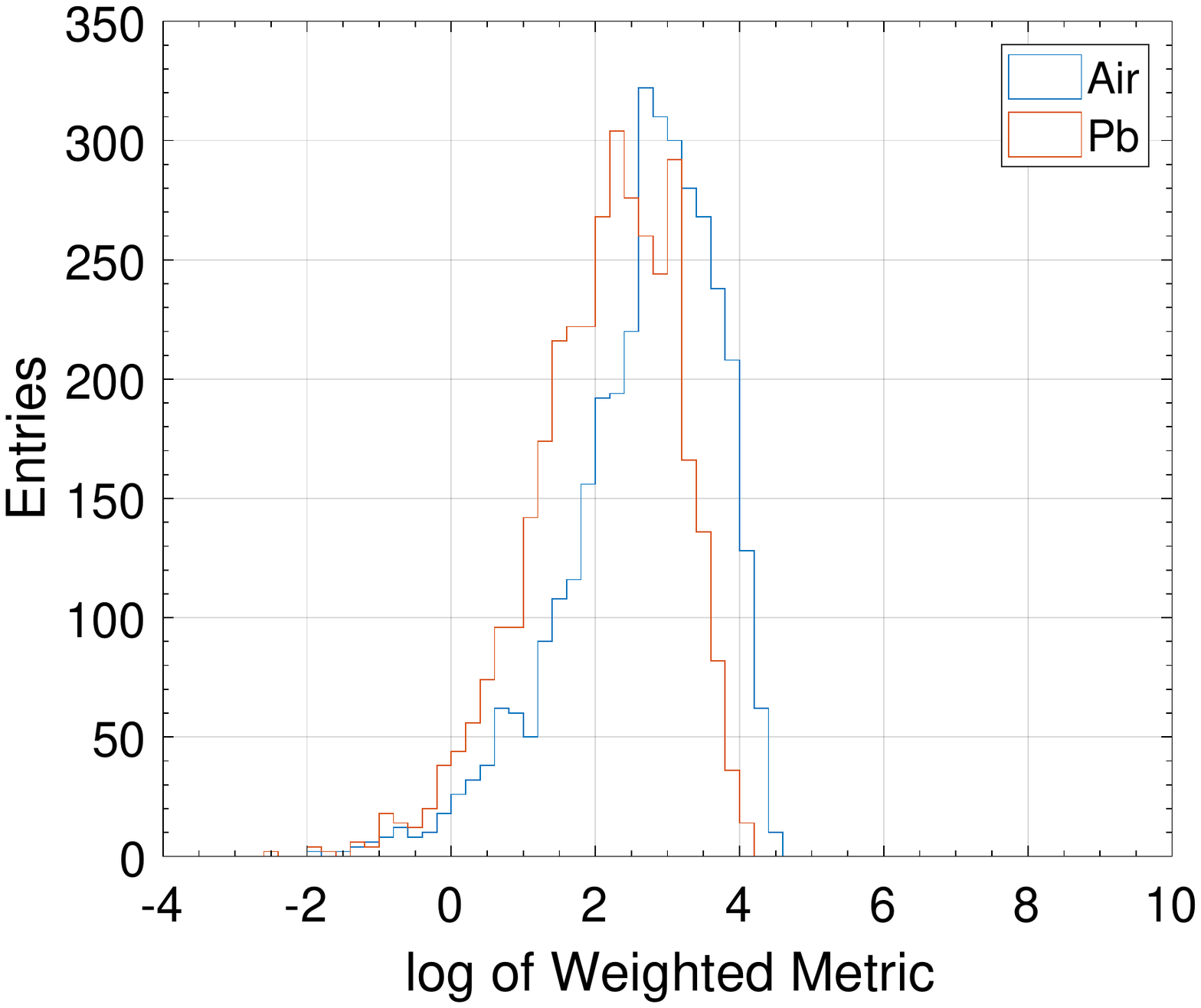}
		\caption{Iterative POCA.}
		\label{IPOCA}
	\end{minipage}
	\hspace*{0.05cm}
	\begin{minipage}[b]{0.32\textwidth}
		\includegraphics[width=\textwidth,height=0.50\textheight,keepaspectratio,trim={1cm 7cm 1cm 7cm},clip]{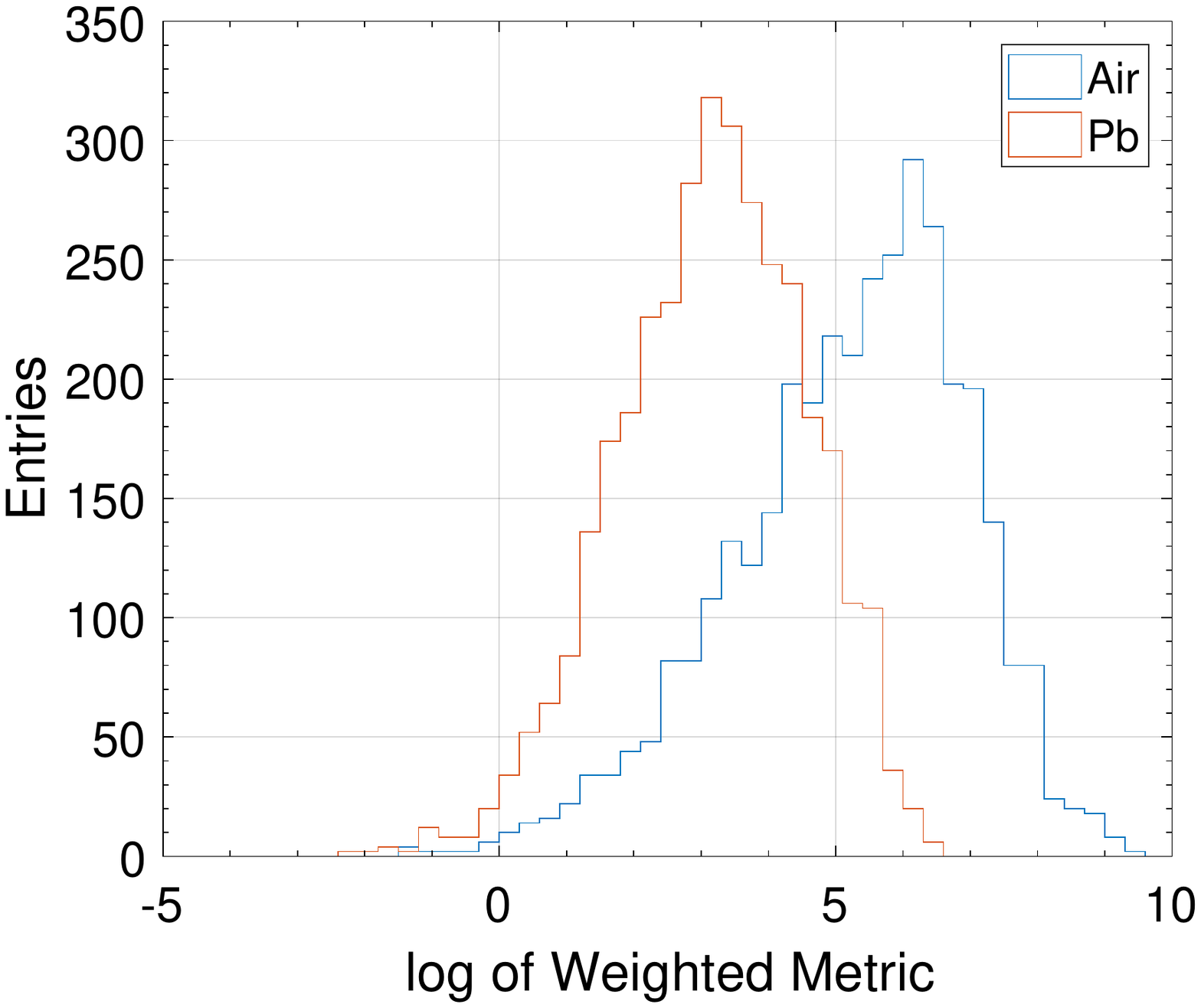}
		\caption{Binned Cluster.}
		\label{VX}
	\end{minipage}
\end{figure}

The distributions of natural logarithm of weighted metric using three different methods have been plotted. The separation of the distributions is a clear indication of the materials present in a sub volume. The metric distance should be less for the sub volumes where target materials are present due to clustering of the vertices. And the weighted metric should be less for the sub volume containing high-$Z$ materials than the ones filled with air. From figures~\ref{POCA} and~\ref{IPOCA}, it can be observed that the algorithms POCA and its iterative variant are able to smell the presence of lead cube of side 5 cm in a sub volume of dimensions 15 $\times$ 15 $\times$ 50 cm${^3}$. However if they are compared to figure~\ref{VX}, it can be stated that its capability of distinguishing the materials is not as good as the binned cluster algorithm, which clearly separates the two distributions by significant amount.
\vskip 3pt
The binned cluster algorithm is then tested to differentiate cubical blocks of side 5 cm in two different sub volumes and result of the observation of 1 hr. data has been shown in figure~\ref{UPb}. Further, the observation of 30 min to distinguish air and lead has been shown figure~\ref{30}. Earlier by the cited group~\cite{bristol} , the algorithm has been reported to be able to differentiate materials in smaller duration. However, in our case the setup is a much smaller prototype, therefore the number of events become significantly low below 30 min time.
\begin{figure}[!tbph]
	\centering
	\begin{minipage}[b]{0.48\textwidth}
		\includegraphics[width=\textwidth,height=0.83\textwidth]{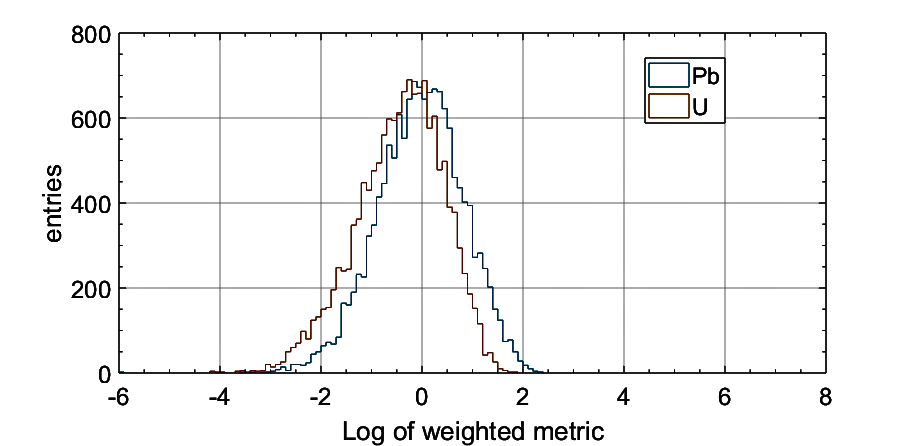}
		\caption{Discrimination of Lead and Uranium. }
		\label{UPb}
	\end{minipage}
	\hspace*{0.1cm}
	\begin{minipage}[b]{0.48\textwidth}
		\includegraphics[width=\textwidth,height=0.45\textheight,keepaspectratio,trim={1cm 6cm 1cm 6cm},clip]{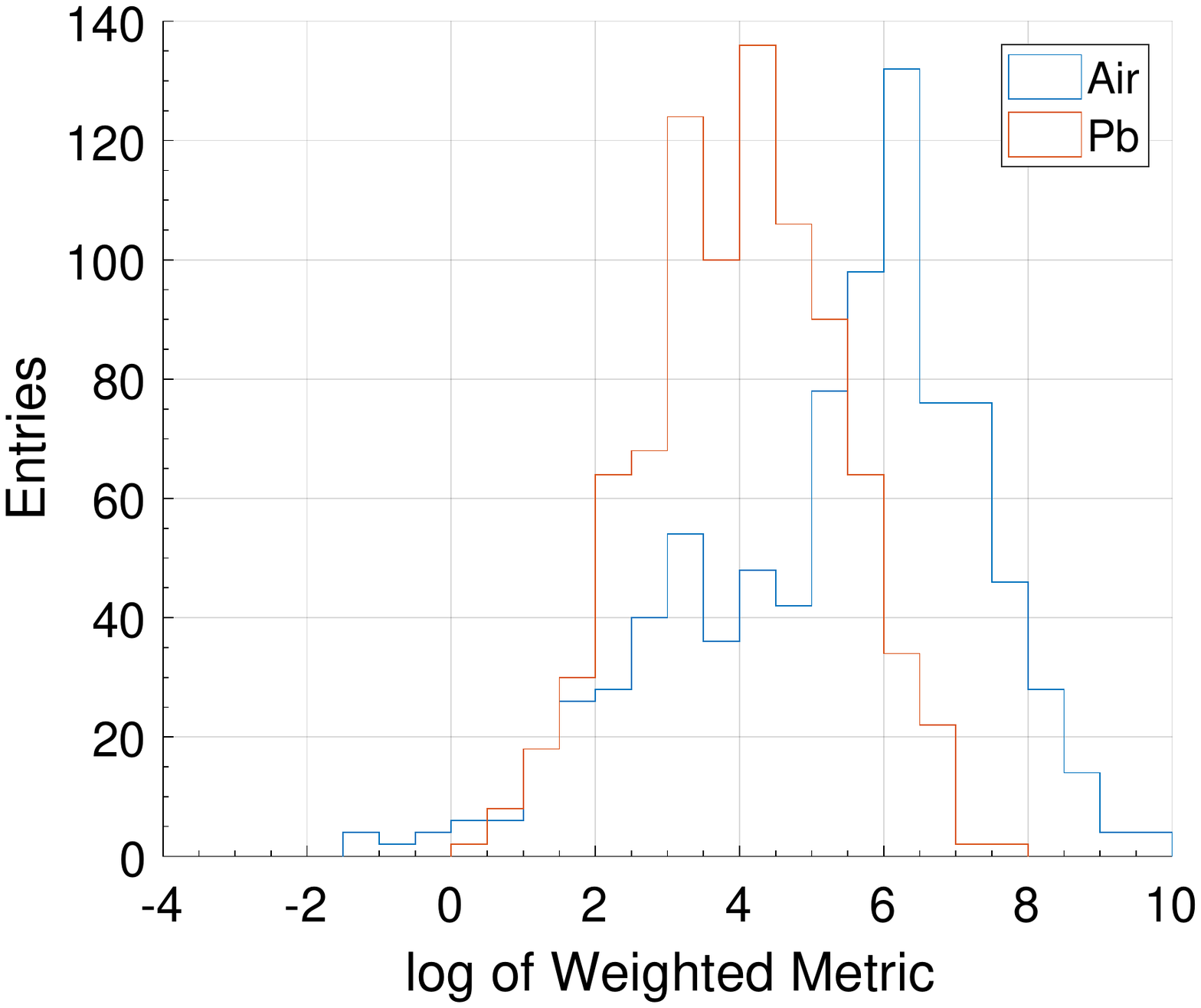}
		\caption{Discrimination of Air from Lead in 30 min.}
		\label{30}
	\end{minipage}
\end{figure}
}

\section{Conclusion}
{
It can be concluded that the scattering angle of muons through material medium is a signature of their $\rho$ and $Z$ of their constituent matter. Although these algorithms described here are based on single point scattering and do not contain the momentum information, these are able to distinguish materials of different~$Z$ values. Even with a bad resolution of 1 cm, the image has been successfully reconstructed by the binned cluster algorithm. All three algorithms have been used to calculate the discriminator which has been able to distinguish lead from air in 1 hr. of cosmic data. The binned cluster algorithm is successful in separating regions with lead from air within 30 min. This algorithm has also been able to differentiate Pb from U in one hour.  With consideration of momenta information of the muons, along with improved spatial resolution and larger area coverage of the detector this algorithm can be useful in material discrimination in shorter time. In future the scope of other algorithms like maximum likelihood algorithm will also be considered for similar investigation. 

}

\acknowledgments
The authors thankfully acknowledge the help and support of the members, scientific and technical staffs of ANP division and Prof. Sudeb Bhattacharya. The author Sridhar Tripathy likes to thank INSPIRE Division, Department of Science and Technologies, Govt. of India for financial assistance.


\end{document}